\documentclass{elsart}
\usepackage{graphicx,harvard,amssymb}
\journal{New Astronomy}

\newcommand{\lsim}{\,\lower2truept\hbox{${<\atop\hbox{\raise4truept\hbox{$\sim$}}}$}\,}
\newcommand{\gsim}{\,\lower2truept\hbox{${>\atop\hbox{\raise4truept\hbox{$\sim$}}}$}\,}

\def\etal   {et~al.\,}

\begin{document}

\begin{frontmatter}

\title{Non-linear evolution of the cosmological 
background density field as diagnostic of the 
cosmological reionization}

\author[Popa]{L.A.~Popa~$^{\rm a,b}$, C.~Burigana~$^{\rm a}$, and N.~Mandolesi} 

\vskip 0.5truecm

\address[Popa]{IASF/CNR, Istituto di Astrofisica Spaziale e Fisica Cosmica,
Sezione di Bologna,\\
Consiglio Nazionale delle Ricerche, Via Gobetti 101, I-40129 Bologna, Italy\\ 
$^{\rm b}$~Institute of Space Sciences, Bucharest-Magurele, R-76900, Romania}


\footnote{The address to which the proofs have to be sent is: \\
Carlo Burigana\\ 
IASF/CNR, Istituto di Astrofisica Spaziale e Fisica Cosmica,
Sezione di Bologna, Consiglio Nazionale delle Ricerche,  
Via Gobetti 101, I-40129 Bologna, Italy\\
fax: +39-051-6398724\\
e-mail: burigana@bo.iasf.cnr.it}

\newpage
\begin{abstract}
We present constraints on the cosmological and reionization
 parameters based on the cumulative mass function of the
 Ly-$\alpha$ systems.
 We evaluate the formation rate of bound objects
 and their cumulative mass function for a class of
 flat cosmological models with cold dark matter
 plus cosmological constant
 or dark energy with constant equation of state encompassing different
 reionization scenarios
 and compare it with the cumulative mass function obtained
 from the Ly-$\alpha$ transmitted flux power spectrum.\\
 We find that the analysis of the cumulative mass function of
 the Ly-$\alpha$ systems indicates a reionization redshift $z_r=24.2\pm4$ (68\%CL)
 in agreement with the value found on the basis of the WMAP
 anisotropy measurements,
 setting constraints on the amplitude of
 the density contrast, $\sigma_8=0.91\pm0.04$ (68\%CL),
 similar to those derived
from the X-ray cluster temperature function.\\

Our joint analysis of  Ly-$\alpha$ cumulative mass function
and WMAP anisotropy measurements
shows that the possible current 
identification of a running of the slope, 
$dn_s/d{\rm ln}k\ne 0$, at $k_p$=0.05Mpc$^{-1}$
(multipole ${\it l} \approx 700$) is
mainly an effect of the existing
degeneracy in the amplitude-slope plane at this scale, the
result being consistent with the absence of running,
the other constraints based on WMAP data remaining substantially 
unchanged.

Finally, we evaluate the progress on 
the determination the considered parameters
achievable by using the final temperature anisotropy data
from WMAP and from the forthcoming {\sc Planck} satellite 
that will significantly improve the sensitivity and reliability
of these results.

This work has been done in the framework of the
{\sc Planck} LFI activities.

\end{abstract}


\end{frontmatter}

\newpage

\section{Introduction}

The cosmological density  background is assumed to have been seeded
at some early epoch in the evolution of the Universe, inflation being the most
popular of the current theories for the origin of the 
cosmological structures (see e.g. Kolb \& Turner 1990, Linde 1990).\\
Measurements of the Cosmic Microwave Background (CMB) 
anisotropies present at the epoch of the recombination
at large scales ($R \gsim 8h^{-1}$Mpc), always outside the horizon
during the radiation-dominated era,
carry precious information on the power spectrum of density field. 
Moving to intermediate and small scales the background
density field encodes information related to
the non-linear evolution of the gravitational clustering,
the time evolution of the galaxy bias
relative to the  underlying mass distribution (see e.g. Hoekstra et al. 2002,
Verde et al. 2002),
and the magnitude of the peculiar motions and bulk
flows in the redshift space (Kaiser 1987).
At these scales the spectrum of the background density field can be
constrained by complementing CMB measurements
with other astronomical data (see Kashlinsky (1998)
for the evaluation of the spectrum of the density field
 from a variety of astronomical data
on scales
1$h^{-1}$Mpc $\le$ R $\le$ 100$h^{-1}$Mpc ).

In particular, on scales less then 5 $h^{-1}$Mpc the spectrum of the density
field can be constrained by using observations of the spatial distribution  
of high-redshift collapsed objects
as quasars and galaxies and information on their macroscopic properties
(Efstathiou \& Ress 1988, Kashlinsky \& Jones 1991,
Kashlinsky 1993). The detection of high-z quasars
by the Sloan Digital
Sky Survey (Becker et al. 2001, White et al. 2003, Fan et al. 2003)
and by the Keck telescope (Vogt et al. 1994, Songaila \& Cowie 2002)
as well as the detection of high-redshift
galaxies (see Kashlinsky (1998) and the references therein)
are indications about  the existence of such early collapsed
objects at redshifts between 2 and 6,  on mass scales 
of $\sim$ 10$^{10}$M$_{\odot}$.\\
The {\it rms} mass fluctuations over a sphere of radius of 8$h^{-1}$Mpc,
$\sigma_8$, fixes the amplitude of the density field power spectrum  
and then determines 
the redshift of collapse, $z_c$, of an object with a given mass-scale.
Consequently, the observation of such objects can set significant constraints
on different cosmological theories.\\
Recently, the WMAP~\footnote{http://lambda.gsfc.nasa.gov}
team (Spergel et al. 2003, Verde et al. 2003)
highlighted the relevance of complementing the
CMB and LSS
measurements with the Ly-$\alpha$ forest data (the absorptions observed in
quasar spectra by the neutral hydrogen in the intergalactic medium)
in constraining the cosmological parameters 
(Croft et al. 1998; Zaldarriaga et al. 2003 and references therein) and  the shape and
amplitude of the primordial density field at small scales.

In this paper we investigate the dynamical effects  of the non-linear
evolution of the density field implied by the high-z collapsed objects
as diagnostic of the reionization history of the Universe.
The reionization is assumed to be caused by the ionizing photons produced
in star-forming galaxies and quasars when the cosmological
gas falls into the potential wells caused by the cold dark matter
halos. 
In this picture the reionization history of the Universe
is a complex process that
depends on the evolution of the background
density field and of the gas properties in the intergalactic medium (IGM)
and on their feedback relation.
The evolution of the background density field, that
determines the formation rate of the bound objects, is a function of the
grow rate of the density perturbations and depends on the assumed
underlying cosmological model.
The evolution of the gas in the IGM is a complex function
of the gas density distribution and
of the gas density-temperature relation.
The latter is related to the spectrum (amplitude and shape)
of the ionizing radiation, to the reionization history
parametrized by some reionization parameters (the
reionization redshift, $z_r$, and the reionization temperature, $T_r$), and
to the assumed cosmological parameters.\\
As shown by hydrodynamical simulations (Cen et al. 1994; Zhang et al. 1995;
Hernquist et al. 1996; Theuns et al. 1998), the gas in the IGM is highly
inhomogeneous, leading to the non-linear collapse of the structures.
In this process the gas is heated to its virial temperature.
The photoionization heating and the expansion
cooling cause the gas density and
temperature to be tightly related.
Finally, the temperature-mass relation for the gas in the IGM at
the time of virilization determines  
the connection between the gas density
and the matter density at the corresponding scales.

Taking advantage of the results of a number of  hydrodynamical 
simulations (Cen et al. 1994; Miralda-Escud\'e et al. 2000; 
Chiu, Fan \& Ostriker 2003)
and semi-analytical models (Gnedin \& Hui 1998;
Miralda-Escud\'e, Haehnelt \& Rees 2000, Chiu \& Ostriker 2000),
we re-assess in this paper the possibility to use  the
mass function of Ly-$\alpha$ systems  to place
constraints on spatially flat cosmological
models with cold dark matter plus cosmological constant or dark energy,
encompassing different reionization scenarios.
We use the Press-Schechter theory (Press \& Schechter 1974)
to compute the  comoving number density
of Ly-$\alpha$ systems
per unit redshift interval
from the Ly-$\alpha$ transmitted flux power spectrum (Croft et al. 2002) and examine the
constraints on the cosmological parameters.
In our analysis we take into account the connection between the non-linear
dynamics of the gravitational collapse and the properties of the gas in the IGM
through the virial temperature-mass relation.
We address the question of the consistency of the WMAP and Ly-$\alpha$ constraints
on the cosmological parameters, paying a particular attention to the degeneracy
between the running of the effective spectral index and
the power spectrum amplitude.
Finally, we discuss the improvements achievable with the final WMAP 
temperature anisotropy data
and, in particular, the impact of the next CMB temperature anisotropy 
measurements on-board the ESA 
{\rm Planck}~\footnote{http://astro.estec.esa.nl/Planck/} satellite.


\newpage

\section{Early object formation mass function}

\subsection{Formation rates}

The most accurate way used to assess the formation rates of the high-z
collapsed objects is based on numerical simulations.
A valid alternative is offered by the  Press-Schechter theory (Press \& Schechter 1974,
Bond et al. 1991) extensively tested by numerical simulations for both open
and flat cosmologies
(Lacey \& Cole 1994, Eke, Cole \& Frenk 1996, Viana \& Liddle 1996).

According to the Press-Schechter theory, the fraction of the mass residing
in gravitationally  bounded objects is given by:
\begin{equation}
f_{{\rm coll}}(z) \approx \sqrt{\frac{2}{\pi}} \frac{1} {\sigma(R_f,z)}
\int_{\delta_c(z)}^{\infty} \exp{\left[-\frac{\delta_c^2(z)}
{2 \sigma(R_f,z)}\right]}
d\,\delta_c(z) \,.
\end{equation}
Here $\delta_c(z)$ is the redshift dependent density threshold
required for the collapse; $R_f$ is the
filtering  scale associated with the mass
scale $M=4 \pi R_f^3 \rho_b/3$,
$\rho_b$ being the comoving background density; 
$\sigma(R_f,z)$ is the
{\it rms} mass fluctuation within the radius $R_f$:
\begin{eqnarray}
\sigma^2(R_f,z)=\int_0^{\infty}\frac{d \,k}{k}\Delta^2(k,z)W^2(kR_f)\, ,
\end{eqnarray}
where W(x) is the window function chosen to filter the density field. For a top-hat
filtering $W(x)=3({\rm sin}x-x{\rm cos}x)/x^3$ while for a Gaussian filtering
$W(x)=\exp(-x^2)$. As we do not find significant differences between the results
derived by adopting the two considered window functions for some
representative cases,
we present in this work the results obtained
by using the Gaussian smoothing. In the above equation $\Delta^2(k,z)$
is referred as the power variance
and is related to the matter power spectrum $P(k,z)$ through:
\begin{equation}
\Delta^2(k,z)=\frac{1}{2 \pi^2}k^3P(k,z)\,.
\end{equation}
Motivated in the framework of the spherical collapse model and calibrated by
N-body numerical simulations, the linear density threshold of
the collapse $\delta_c(z)$
was found to vary at most by $\simeq 5$~\% with the background cosmology
(see e.g. Lilje 1992, Lacey \& Cole 1993,
Eke, Cole \& Frenk 1996). However,  the choice of $\delta_c(z)$ depends on
the type of collapse. For the spherical collapse the standard choice
is $\delta_c(0)=1.7 \pm 0.1$ while for pancake formation or filament formation
its value is significantly smaller (Monaco 1995). For the purpose of this work
we assume that the collapse have occurred spherically and use $\delta_c(0)=1.686$
that is the conventional choice for the case of the flat cosmological models
(Eke, Cole \& Frenk 1996).
The time evolution of $\delta_c$ depends on the background cosmology:
\begin{equation}
\delta_c(z)=\delta_c(0)\frac{D(z)}{D(0)} \, ,
\end{equation}
where $D(z)$ is the linear growth function of the density perturbation, given in 
general form by (Heath 1977, Carroll, Press \& Turner 1992, Hamilton 2001):
\begin{equation}
D(a)=\frac{5\Omega_m}{2 a f(a)} \int_0^a f^3(a) d \,a \, .
\end{equation}
Here $a=(1+z)^{-1}$ is the cosmological scale factor normalized to unity
at the present time ($a_0=1$), $\Omega_m$ is the matter density energy parameter
at the present time and $f(a)$ specifies  the time evolution of the
scale factor for a given cosmological model:
\begin{eqnarray}
\frac{d\,a}{d \,t}=\frac{H_0}{f(a)}, \hspace{0.4cm}
f(a)=\left[ 1+\Omega_m\left(\frac{1}{a}-1\right) +
\Omega_{de}\left( \frac{1}{a^{1+3w}}-1\right) \right]^{-1/2}.
\end{eqnarray}
In the above equation $H_0$ is the present value of the Hubble paramenter,
$\Omega_{de}$ is the present value of the energy density parameter of the
dark energy, $w=p/\rho_{de} \sim -1$ defines the dark energy
equation of state, and $\Omega_m$ is the matter energy density parameter.
Equation (6) reduces to that of a $\Lambda$CDM model for $w=-1$.

The comoving number density of the gravitationally collapsed objects
within the mass interval $ d\,M$ about $M$ at a redshift $z$
is given by (Viana \& Liddle 1996):
\begin{equation}
n(M,z)d\,M=-\sqrt{\frac{2}{\pi}}\frac{\rho_b}{M}\frac{\delta_c}{\sigma^2(R_f,z)}
\frac{d\,\sigma(R_f,z)}{d\,M}
\exp{\left[-\frac{\delta_c^2}{2 \sigma^2(R_f,z)}\right]}d\,M.
\end{equation}
We are interested in the formation rate of the high-z collapsed
objects at a given redshift.
According to Sasaki method  (Sasaki 1994), the comoving number density of
bounded objects with the mass in the range $dM$ about M, which virilized
in the redshift interval $dz$ about $z$ and survived until the redshift
$z_f$ without merging with other systems is given by:
\begin{equation}
 {\rm N }(M,z)d\,M d\,z=\left[
 -\frac{\delta_c^2}{\sigma^2(R_f,z)}
         \frac{n(M,z)}   {\sigma(R_f,z)}
         \frac{d\, \sigma(R_f,z)}{d\,z}\right]
          \frac{\sigma(R_f,z)}{\sigma(R_f,z_f)}
          d\,M \,d\,z\, ,
\end{equation}
where: $$\sigma(R_f,z)=\sigma(R_f,0)\frac{D(z)}{D(0)}\frac{1}{1+z}.$$
The total comoving number density of the bounded objects per unit redshift
interval with the mass exceeding $M$ (the cumulative mass function) is given by:
\begin{equation}
{\bf N}(>M)=\int^{\infty}_M {\rm N}(M,z)dM.
\end{equation}

\subsection{Mass-temperature relation for the virilized gas in the IGM}

The fraction of the mass of the gas in collapsed virilized halos
can be  calculated
if the probability distribution function (PDF)
for the gas overdensity is known:
\begin{eqnarray}
f_{{\rm coll}}(z)=\int_{\Delta_c(z)}^\infty {\bf \Delta} P_V({\bf \Delta})
d\,{\bf \Delta}\,.
\end{eqnarray}
Here $P_V({\bf \Delta})$ is the volume-weighted PDF
for the gas overdensity ${\bf \Delta}=\rho_g/\rho_{bar}$, where $\rho_g$ is the
gas density and $\rho_{bar}$ is the mean density of baryons.
In the above equation $\Delta_c(z)$ is the halo density contrast at virilization.
Based on hydrodynamical simulations,
Miralda-Escud\'e et al. (2000) found for the volume-weighted
probability distribution, $P_V({\bf \Delta})$, the following fitting
formula:
\begin{equation}
P_V({\bf \Delta)}d\, {\bf \Delta}=A \exp{\left[-\frac{({\bf \Delta}^{-2/3}-C_0)^2}
{2(2\delta_0/3)^2}\right]}{\bf \Delta}^{-\beta} d \, {\bf \Delta}.
\end{equation}
In this equation  $\delta_0$ is the linear {\it rms} gas
density fluctuation and $\beta$ is a parameter that describe the
gas density profile ( $\rho_g \sim r^{-\beta}$ for an isotermal gas).
As shown by the numerical
simulations (see Table 1 from Chiu, Fan \& Ostriker 2003)
the redshift evolution  of $\delta_0$
depends on the underlying cosmological model.
Assuming the same fraction of baryons and dark matter
in collapsed objects,
we compute the redshift dependence of $\delta_0$
on the cosmological parameters
by using an iterative procedure 
(Chiu, Fan \& Ostriker 2003)
requiring the equality of
the equations (1) and (10) 
representing the collapsed mass fraction.
The parameters $A$ and $C_0$
were obtained by requiring the normalization to unity
of the total volume and mass.
For the redshift dependence
of $\beta$ in the considered redshift range
we take the values 
$\beta \approx {\rm min}[2.5, \, 3.2-4.73/(1+z)]$
obtained by Chiu, Fan \& Ostriker (2003) 
through a fit to their hydrodinamical  simulations.

As we are interested to apply equation (9) to the virilized gas,
we need to know the temperature-density and the mass-temperature relations
for the virilized gas in the IGM. \\
The temperature-density relation is determined by the reionization
scenario and the underlying cosmological model.
Hydrodynamical simulations can predict this relation accurately,
but the limited computer resources restrict the number of cosmological
models and reionization histories that can be studied.
For this reason we evaluate the temperature-density relation at the redshifts
of interest by using the semi-analytical model developed by Hui \& Gnedin (1997)
that permits to study the reionization models by varying the amplitude,
spectrum, the epoch of reionization and the underlying cosmological
model. According to this model,
for the case of uniform reionization models,
the mean temperature-density relation is well approximated  
by a power-law equation of state that can be written as:
\begin{equation}
T=T_0(1+{\bf \Delta})^{\gamma-1},
\end{equation}
where $T_0$ and $\gamma$ are analytically computed as  functions of
the reionization
temperature $T_r$, the reionization redshift $z_r$, the matter and baryon 
energy density parameters $\Omega_m$ and $\Omega_{bar}$, 
and the Hubble parameter $H_0$.

According to the virial theorem (Lahav et al. 1991, Lilje et al. 1992),
the virial mass-temperature relation at any redshift $z$ can be written
as (Eke, Cole \& Frenk 1996; Viana \& Liddle 1996; Kitayama \& Suto 1997;
 Wang \& Steinhardt 1988):
\begin{equation}
\frac{M_{vir}}{10^{15} h^{-1} M_{\odot}}=\left(\frac{k_BT_{vir}/{\it f}_{\beta}}{0.944{\rm keV}}\right)^{3/2}
[(1+z)^3 \Omega_0 \Delta_c]^{-1/2}
\left[1-\frac{2 \Omega_{de}(z)}{\Delta_c \Omega_m(z)}\right]^{-3/2}  ,
\end{equation}
where $k_B$ is the Boltzmann constant, $T_{vir}$ is the temperature of
the virilized gas,
$\Delta_c$ is the density contrast
at virilization and  ${\it f}_{\beta}= {\it f}_u \mu/\beta$,
where ${\it f}_u$ is the fudge factor (of order of unity) that allows
for deviations from the
simplistic spherical model and $\mu$ is the proton molecular weight.
Different analyses adopting similar mass-temperature relations
 disagree on the value of $f_{\beta}$
 because of the uncertainties in the numerical simulations.
 We adopt here $f_{\beta}=1$,  as indicated by
 the most extensive simulation results obtained by Eke, Cole \& Frenk (1996).

 Throughout this paper we consider that the virilization takes place at
 the collapse time, $t(z_c)$,  that is half of the turn-around time:
 $t(z_c)=t(z_{ta})/2$, $z_{ta}$ being the redshift at which
 ${\dot R}(z_{ta})=0$ ($R$ is the radius of a
 spherical overdensity).\\
 For $\Lambda$CDM models the density contrast at virilization
 is a function of $\Omega_m$ only.
  For quintessence  models the
 density contrast at virilization becomes a function
 of $\Omega_m$ and $w$ and can be written as
 (Wang \& Steinhardt 1988):
\begin{equation}
\Delta_c(z=z_c)=\frac{\rho_{clust}(z_c)}{\rho_b(z_c)}=
\zeta\left(\frac{R_{ta}}{R_{vir}}\right)^3
\left(\frac{1+z_{ta}}{1+z_c}\right)^3,
\end{equation}
where:
$\zeta(z_{z_{ta}})=
\rho_{clust}(z_{ta})/\rho_b(z_{ta})$; $R_{ta}$ and $R_{vir}$
are the radius at $z_{ta}$ and $z_c$ respectively:
\begin{equation}
\zeta(z_{ta})=(3 \pi/4)^2\Omega_m(z_{ta})
^{-0.79+0.26 \Omega_m(z_{ta})-0.06w},
\end{equation}
\begin{equation}
\frac{R_{vir}}{R_{ta}}=\frac{1-\eta_v/2}{2+\eta_t-3\eta_v/2},
\end{equation}
where
$\eta_t=2 \zeta^{-1}\Omega_{de}(z_{ta})/\Omega_m(z_{ta})$
and
$\eta_v=2 \zeta^{-1}[(1+z_c)/(1+z_{ta})]^3
\Omega_{de}(z_c)/\Omega_m(z_c)$. \\
The energy density parameters and the Hubble parameter
evolve with the scale factor according to:
\begin{eqnarray}
\Omega_m(a)=\frac{\Omega_0f^2(a)}{a}, \hspace{0.3cm}
\Omega_{de}(a)=\frac{\Omega_{de}f^2(a)}{a^{1+3w}},
\hspace{0.3cm} H(a)=\frac{H_0}{a f(a)},
\end{eqnarray}
where $f(a)$ is given by the equation (6).\\
The  scale factors for collapse, $a_c$, and turn-around, $a_{ta}$,
was computed from the spherical collapse model
(Lahav et al. 1991, Eke, Cole \& Frenk 1996).\\
For any region inside the radius $R$
that was overdense by $\Delta_i$
with respect to the background at some initial time $t_i$
corresponding to the redshift $z_i$, 
we solve the set of equations:
\begin{eqnarray}
\int^{a_{ta}}_0 f(a)d\,a= \frac{H_0} {H_i} \int_0^{s_{ta}}g(s) d\,s,
\hspace{0.3cm}
\int^{a_{c}}_0 f(a)d\,a= 2\frac{H_0} {H_i} \int_0^{s_{ta}}g(s) d\,s,
\end{eqnarray}
where:
\begin{eqnarray}
\frac{ds}{dt}=\frac{H_i}{g(s)} \hspace{0.2cm} {\rm and} \hspace{0.2cm}
g(s)=\left[1+\Omega_i(1+\Delta_i)\left( \frac{1}{s}-1\right)+\Omega_{de,i}
(s^2-1)\right]^{-1/2}.
\end{eqnarray}
In the above equations,
solved by using an iterative procedure, 
$s=R/R_i$ is the scale factor of a spherical perturbation
with the initial radius $R_i$ and $s_{ta}=R_{ta}/R_i$ is its scale factor
at the turn-around (see Appendix A in Eke, Cole \& Frenk 1996).
The average density perturbation inside the radius $R$ is given by:
\begin{equation}
\Delta(R,z)=\frac{3}{R^3}\int_0^R R^2 \delta(R,z) d\,R,
\end{equation}
where $\delta(R,z)$ is related to the power spectrum of the density field
$P(k,z)=|\delta_k(z)|^2$ through:
\begin{equation}
\delta(R,z)=\frac{1}{(2 \pi)^3}\int\delta_k(z) e^{-ikR}d^3\,k.
\end{equation}
As initial conditions for the spherical infall we choose the epoch given by
$z_i=1100$ when the growth of perturbations is fully determined by the linear theory.
The initial values of  the parameters
$H_i=H(z_i)$, $\Omega_i=\Omega(z_i)$ and $\Omega_{de,i}=\Omega_{de}(z_i)$
are given by the equation (17) and for the normalization of the density
field at $z_i$, $\sigma_8(z_i)$, we take:
$$\sigma_8(z_i)=\sigma_8(0)\frac{D(z_i)}{D(0)}\frac{1}{1+z_i}.$$
We adopt for $\sigma_8$ at the present time the value
obtained from the analysis of the local X-ray temperature
function for flat cosmological models
with a mixture of cold dark matter and cosmological constant
or dark energy with constant equation of state (Wang \& Steinhardt 1998):
\begin{eqnarray}
\sigma_8=(0.50-0.1\Theta)\Omega_m^{-\gamma(\Omega_m,\Theta)}\, ,
\end{eqnarray}
where:

\begin{eqnarray}
\gamma(\Omega_m,\Theta)=0.21 -0.22w+0.33\Omega_m+0.25\Theta \nonumber
\end{eqnarray}
and
\begin{eqnarray}
\Theta=(n_s-1)+(h-0.65).  \nonumber
\end{eqnarray}

\section{Results}


\subsection{Cosmological constraints from Ly-$\alpha$ observations}

The study Ly-$\alpha$ transmitted flux  power spectrum
has become increasingly important for cosmology as it is
probing the absorptions produced by the low density gas in voids or mildly
overdense regions. This gas represents an accurate
tracer of the distribution of the dark matter at
the early stages of the structure formation.
One of the most important application
is to recover the linear matter power spectrum $P_L(k)$
from the flux power spectrum  $P_F(k)$ and inferring the cosmological
parameters of the underlying cosmological model. \\
On the observational side there are recent analyses by
McDonald et al. (2000) and
Croft et al. (2002)
that obtain results for the transmitted
flux power spectrum $P_F(k)$ in agreement with each other within the error bars.
Two different methods have been proposed to constrain
the cosmological parameters:
McDonald et al. (2000) and Zaldarriaga et al. (2001) directly compare $P_F(k)$
with the predictions of the cosmological models, while Croft et al. (2002)
and Gnedin \& Hamilton (2002) use an analytical fitting function to recover
the matter power spectrum, $P_L(k)$, from the flux power spectrum $P_F(k)$.
The WMAP team (Verde et al. 2003) used the analytical fitting function
obtained by Gnedin \& Hamilton (2002) to convert $P_F(k)$ into $P_L(k)$.\\
In a recent work, Seljak, McDonald \& Makarov (2003)
investigate the  cosmological implications of the conversion
between the measured flux power spectrum and the matter power
spectrum, pointing out several issues that lead to the expansion
of the errors on the inferred cosmological parameters.

We compute the total comoving number density, $N_{Ly-\alpha}$, 
of Ly-$\alpha$ systems per unit redshift interval that survived until ${z}=2.72$ without
merging with other systems
from the flux transmission
power spectrum, $P_F(k)$, obtained by
Croft at al. (2002) for their fiducial sample with
mean absorption redshift ${\bar z}=2.72$.
We then compare $N_{Ly-\alpha}$ with the theoretical predictions for the
same function, $N_{th}$, obtained for a class of cosmological
models encompassing the dark energy contribution with  constant equation of state
and different reionization scenarios. \\
Our fiducial background cosmology is described by
a flat cosmological model, $\Omega_0=\Omega_m +\Omega_{de}=1$,
with the following parameters at the present time:
$\Omega_{bar}h^2=0.024\pm 0.001$, $\Omega_mh^2=0.14\pm0.02$, $h=0.72\pm0.05$,
as indicated  by the best fit of the power law $\Lambda$CDM model of WMAP data
(Spergel et al. 2003).
In our analysis we allow to vary the primordial scalar spectral index
$n_s$, the parameter $w$ describing the equation of state for the dark energy, 
the reionization redshift $z_r$, and the reionization temperature $T_r$. 
We assume adiabatic initial conditions
and neglect the contribution of the tensorial modes.
\begin{figure}
\caption{The dependence of the virilized gas properties
on the cosmological and  reionization parameters:
$n_s=0.99$, $z_r=10$, $w=-1$, $T_{r,4}=2.5$ (solid lines),
$n_s=0.99$, $z_r=17$, $w=-1$, $T_{r,4}=2.5$ (dashed lines),
$n_s=0.99$, $z_r=10$, $w=-1$, $T_{r,4}=2$ (dot-dashed lines),
$n_s=1.1$,  $z_r=10$, $w=-1$, $T_{r,4}=2.5$ (small dot-dashed lines),
$n_s=1.1$,  $z_r=10$, $w=-0.7$, $T_{r,4}=2.5$ (dotted lines).
$T_{r,4}$ represents the reionization temperature in units of $10^4$K.
See also the text.}
\begin{center}
\includegraphics[width=16cm]{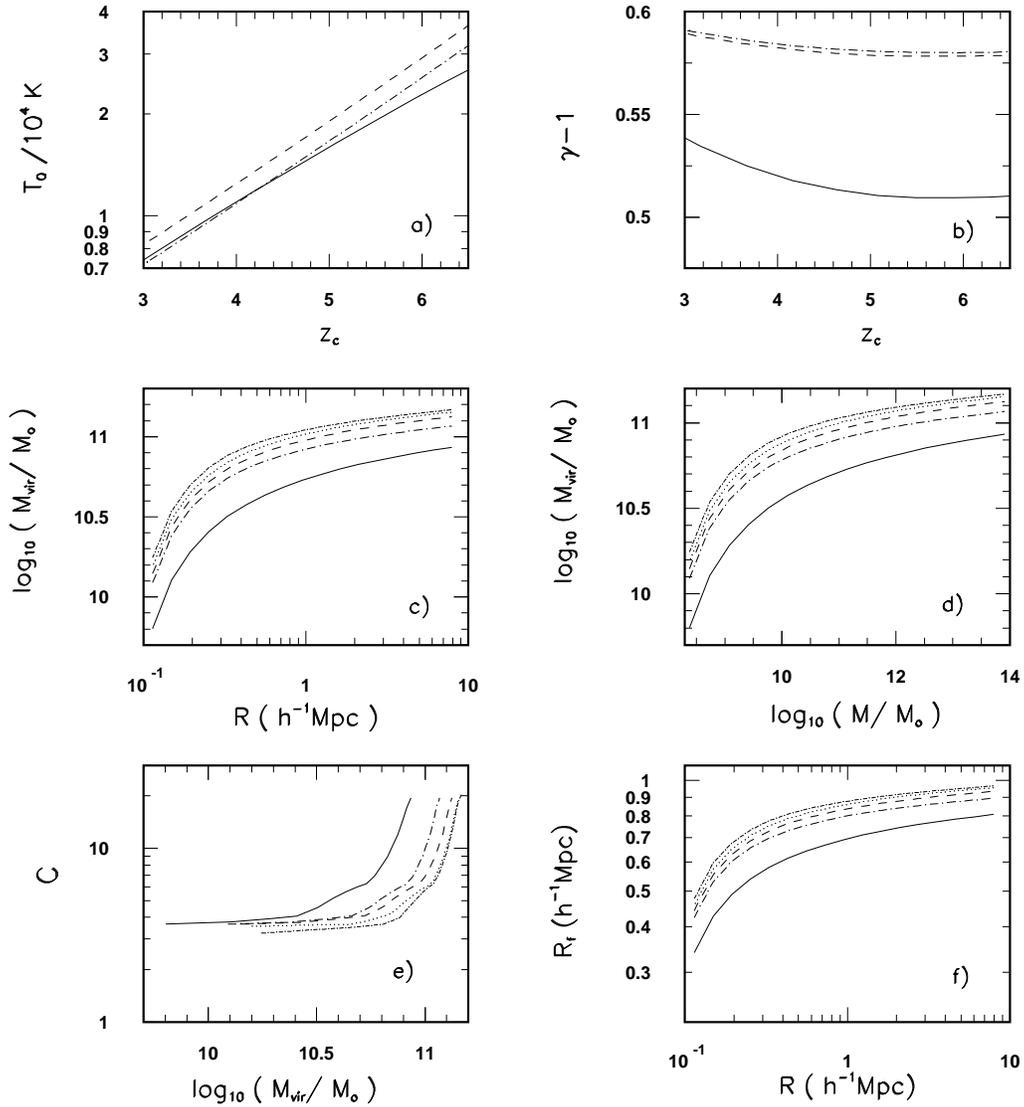}
\end{center}
\label{}
\end{figure}

Our parameter vector ${\bf p}=(n_s,w,z_r,T_r)$ has four dimensions. We create
a grid of model predictions for the
each choice of the parameters in the grid:
\begin{itemize}
\item $n_s=(0.7, 0.8, 0.85, 0.99, 1.1, 1.15, 1.2, 1.3)$
\item $w=(-0.4, -0.5, -0.6, -0.65, -0.7, -0.75, -0.8, -0.85, -0.9, -0.95, -1)$
\item $z_r=(5, 6, 8, 10, 12, 14, 16, 18, 20, 22, 24, 26, 28, 30, 35, 40)$
\item $T_{r,4}=(1.6, 1.8, 1.9, 2, 2.1, 2.2, 2.3, 2.4, 2.5, 2.6, 2.8, 3)$
\end{itemize}
Here $T_{r,4}$ is the reionization temperature in units of $10^4$K.\\
The density perturbations $\delta_k(z_i)$ at the initial redshift $z_i=1100$
and the matter transfer function
at $z=2.72$ was computed for each set of parameters in the grid
by using the CMBFAST code version 4.2 (Seljak \& Zaldarriaga 1996).
Then we evaluate the linear matter power spectrum, $P_L(k)$, with
the appropriate normalization.\\
The Ly-$\alpha$ transmission power spectrum for the fiducial sample
at ${\bar z}=2.72$ proves linear scales in the range $R = (11.4 - 1750)$~km/s
(see Table~2 from Croft et al. 2002).
For the purpose of this work we consider scales up to 791~km/s,
 which are non-linear today.
We compute the averaged density perturbation $\Delta_i$ inside
each scale $R$ at the initial redshift $z_i$ and the corresponding redshift
of collapse, $z_c$, as described  in the previous section.
Then, assuming that the virilization takes
place at the collapse time, we
evaluate the temperature-density relation at $z_c$
and compute the appropriate virial mass at $z=2.72$ by using
the mass-temperature relation.
The virial mass obtained in this way  is related to the filtering
scale through $M_{vir}=(4 \pi / 3) R_f^3 \rho_b({\bar z})$ with
$ \rho_b({\bar z})=(3H_0^2 / 8 \pi G) \Omega_0(1+{\bar z})^3$.
We found that the filtering scale obtained in this way, 
$R_f \approx R_J/2$, 
is proportional to the
Jeans scale, $R_J$, corresponding to the  
the Jeans mass
$M_J= 1.5 \, {\rm T}_4 (1+z_c)^{-3/2}\Omega_m^{-1/2} 10^{10}{\rm h}^{-1}
{\rm M}_{\odot}\,$ ($T_4$ being the gas temperature in units of $10^4$~K),
but the exact relation depends on the cosmological model and
reionization parameters.\\
Figure 1 presents few dependences of the gas properties on the
background cosmology and reionization parameters.
Panels a) and b) show the
dependence of the gas temperature $T_0$ and of the parameter $\gamma$
on the redshift of collapse $z_c$.
Panel c) and d) present the dependence of the virilized mass
on the linear scale $R$ and on the mass scale $M(R)$,
as an indication
on the fraction of the virilized mass at the given scale.
In panel e) we show the
dependence of the clumping factor ${\it C}=<\rho^2_g>/< \rho_{bar}>^2$ on the
virial mass. Panel f) presents the dependence of the filtering scale $R_f$
on the linear scale $R$.\\
We note that the filtering scale obtained in this way
depends on our parameter vector:
$R_f=R_f(k,{\bf p})$, where $k$ is the wave number corresponding
to the linear scale $R$.       \\
For each choice of the parameters in our simulation grid we compute
 $N_{Ly-\alpha}$ according to the equations (7)~--~(9) by
 filtering the transmission power spectrum $P_F(k)$ at each wave number $k$ with
 the corresponding filtering scale $R_f(k,{\bf p})$ and
 apply the same procedure to compute $N_{th}$ from the linear power spectrum
 $P_L(k)$. We compare $ N_{Ly-\alpha}$ and $N_{th}$ by computing
  a Gaussian approximation of the likelihood
 function:
 \begin{equation}
 {\it L}(N_{Ly-\alpha};{\bf p}) \propto \prod_{i=1}^{17}=\exp\left[-\frac{1}{2}
 \left( \frac{N^i_{Ly-\alpha}-N^i_{th}}{\sigma_i}\right)^2\right] \, ,
 \end{equation}
 where $i$ runs over different points [we are using first 17 bins from the
 $P_F(k)$] and $\sigma_i$ is the error bar
 associated to $N_{Ly-\alpha}$ on each point.
 The full chi-squared goodness of fit is $\chi^2 \approx -2 $ln$ {\it L}$ with
 12 degrees of freedom (ndf). We define the confidence level (CL) as the upper
 tail probability of the chi-squared distribution and calculate
 $\chi^2({\rm CL},{\rm ndf})$ for a given CL by
 inverting  the chi-squared distribution. \\
 \begin{figure}
\caption{The cumulative mass function of the Ly-${\alpha}$ systems
per unit redshift interval at $z=2.72$ (filled and open circles) compared
with the corresponding theoretical predictions (solid and dashed lines).
See also the text.}
\begin{center}
\includegraphics[width=15cm]{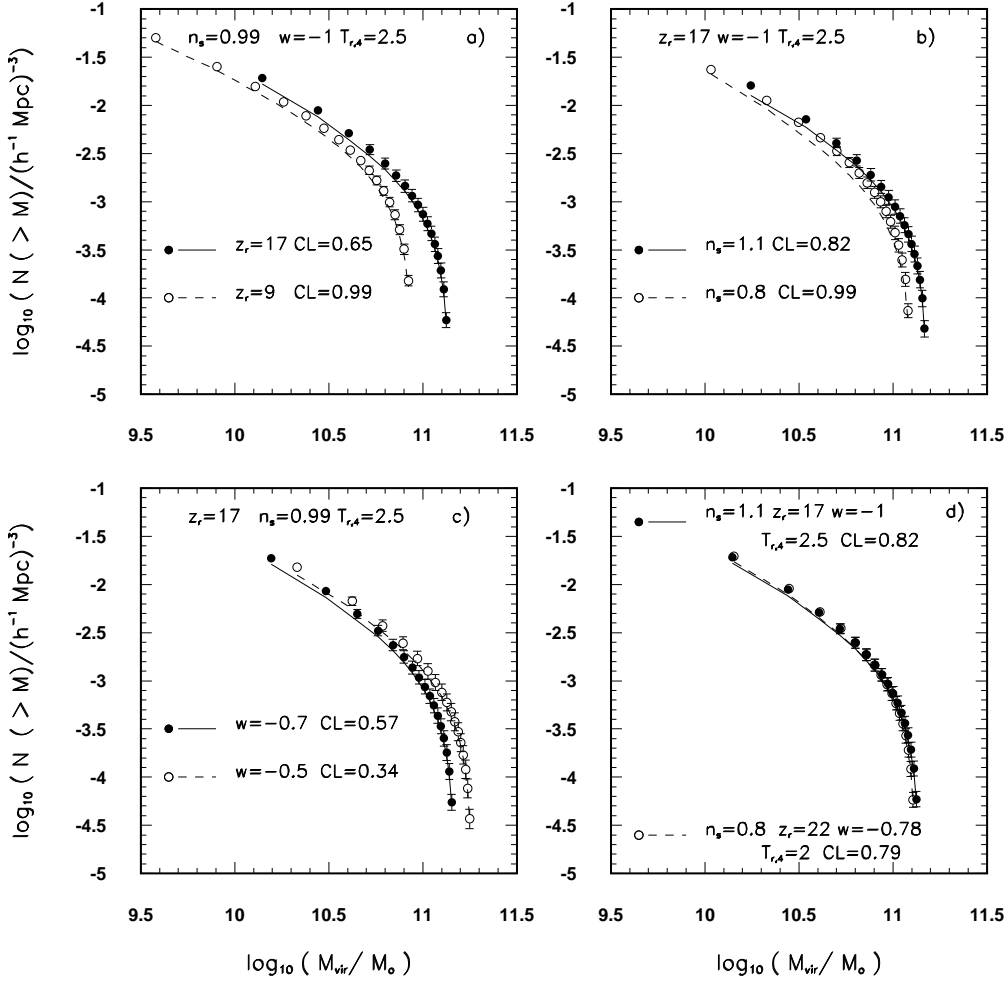}
\end{center}
\label{}
\end{figure}
Figure 2 presents $N_{Ly-\alpha}$ and $N_{th}$ functions obtained for few choices
of the parameters in the simulation grid.
In each  of the panels a), b),
and c) we indicate in the top 
the values of the parameters in common to the two curves
reported in each of the panels a), b),
and c) are indicated in the top while 
the values of the parameters
associated only to one of the two curves
of each of the panels a), b), and c) are indicated in the bottom
together with the corresponding confidence level obtained
from our analysis.
Panel d) shows an example of two quasi-degenerated models and their predictions
for $N_{Ly-\alpha}$.

Each simulated $P_L(k)$ at $z=2.72$ was parametrized
by its effective slope (Peacock \& Dodds 1996), 
the running of the slope and the power variance
at the pivot point $k^*_p$=0.03 s/km:
\begin{equation}
n^*_{{\rm eff}}(k^*_p)=\frac{d {\rm ln}P_L(k)}{d{\rm ln}k} (k=k^*_p/2) \, ,
\end{equation}
\begin{equation}
\frac{d n^*_{\rm eff}}{d {\rm ln} k}(k=k^*_p/2) \, ,
\end{equation}
\begin{equation}
\Delta^2_*(k^*_p)=2 \pi^2 k^3 P_L(k)|_{k^*_p} \, .
\end{equation}

Assuming full ionization 
(ionization fraction $x_e$=1)
for an easier comparison with the 
results of the WMAP team,
we compute for each model in our grid 
the optical depth to the
last scattering, $\tau$, 
and the  normalization of the matter power spectrum
at the present time in terms of $\sigma_{8}$.     \\
Once we have computed the value of the $\chi^2$ for every choice of the parameters
of our simulation grid we marginalize along one direction at a time to get
the four-dimensional constraints on the considered parameters.

\newpage
In Table 1 we report the 
constraints~\footnote{We also verified that 
the final results do not change significantly 
by using the transmission power spectra from 
McDonald \& Miralda-Escud\`e (1999) at $z = 2.4, 3$ and 3.9.} 
we have obtained at
$68\%$ CL and $95\%$ CL. 
Figure 3 presents our $68\%$ CL contour
in $n^*_{{\rm eff}}$ - $\Delta^2_*$ plane, indicating  the best fit values and
in Figure 4 we compare the  best fit linear power
spectrum, $P_L(k)$, at $z=2.72$  ($68\%$ CL)
with the transmission power spectrum $P_F(k)$.
The error bars  include the statistical
error and a systematic error due to the
undeterminations in  $\Omega_m$,
$\Omega_{de}$ and $H_0$.\\
Our values for  $n^*_{{\rm eff}}$
and  $\Delta^2_*$  are in a good agreement with
those obtained by Seljak, McDonald \& Makarov (2003)  indicating
the same degeneracy direction
in $n^*_{{\rm eff}}$ - $\Delta^2_*$ plane,
but are only marginally consistent with the similar values obtained
by Croft et al. (2002).   \\
 We find that the analysis of the cumulative mass function of
 the Ly-$\alpha$ systems indicates a reionization redshift 
 in agreement, within the error bars, 
 with the value found on the basis of the WMAP
 anisotropy measurements,
 setting constraints on the amplitude of
 the density contrast
 similar to those derived
from the X-ray cluster temperature function.
\begin{table}[]
\caption[] {Cosmological parameter constraints from Ly-$\alpha$
 cumulative mass function.
The values of  $n^*_{{\rm eff}}$, $d n^*_{{\rm eff}}/d {\rm ln}k$ and $\Delta^2_*$
are evaluated at the pivot point $k^*_p$=0.03 s/km and  $z=2.72$;
$T_{r,4}$ is the reionization temperature
in units of $10^4$K. }
\begin{center}
\begin{tabular}{ccc}
\hline \hline
                               & Ly-$\alpha$&         \\ \hline
Parameter                      &$68\%$ CL & $95\%$ CL            \\ \hline
$n_s$                          & 1.001$\pm$0.034&0.975$\pm$0.047      \\
$T_{r,4}$                      & 2.317$\pm$0.205& 2.244$\pm$0.227      \\
$z_r$                          & 24.195$\pm$3.976&22.313$\pm$4.814             \\
$w$                            & $-$0.689$\pm$0.141 &$-$0.729$\pm$0.144  \\
$\sigma_8$                     & 0.911$\pm$0.038&0.919$\pm$0.039 \\
$\tau^a$                       & 0.148$\pm$0.035&0.132$\pm$0.041   \\
$\sigma_8 e^{-\tau}$           &0.786$\pm$0.041&0.805$\pm$0.047     \\
$n^*_{{\rm eff}}$              &$-$2.550$\pm$0.034&$-$2.576$\pm$0.047\\
$d n^*_{{\rm eff}}/d {\rm ln}k$ &$-$0.017$\pm$0.004&$-$0.018$\pm$0.005\\
$\Delta^2_*$                   &0.666$\pm$0.113&0.618$\pm$0.131   \\ \hline
$^a$Assumes ionization fraction, $x_e$=1.
\end{tabular}
\end{center}
\end{table}
\begin{figure}
\caption{Constraints at $68\%$ CL  on the effective slope $n^*_{{\rm eff}}$
and power variance  $\Delta^2_*$ at $k^*_p$=0.03 s/km and  $z=2.72$
from:  Ly-$\alpha$ cumulative mass function analysis
(thin solid contour and filled square),
WMAP anisotropy measurements (dashed contour and filled circle)
and the joint WMAP and Ly-$\alpha$
analysis (thick solid --~green~-- contour and open circle).
See also the discussion in Sect.~3.2.
The similar values obtained by Croft et al. 2002 (filled triangle)
are also indicated.}
\begin{center}
\includegraphics[width=13cm]{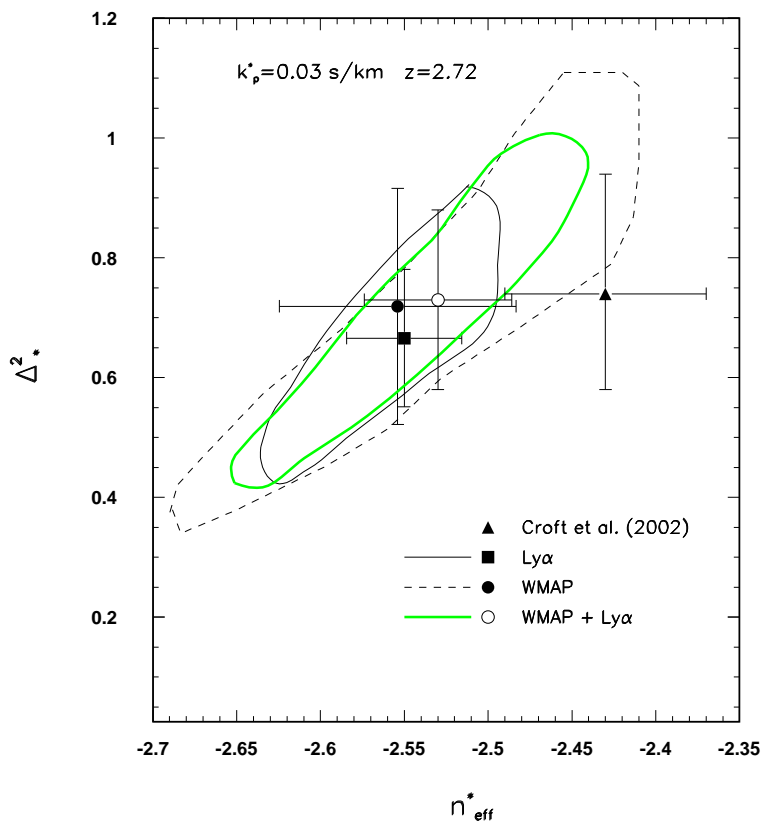}
\end{center}
\label{}
\end{figure}
\begin{figure}
\caption{The best fit linear power spectrum $P_L(k)$ at $z=2.72$
 obtained at 68\% CL from the analysis of the
Ly-$\alpha$ cumulative mass function (open circles and solid line)
compared with the transmission
power spectrum $P_F(k)$ from Croft et al. 2002 (filled circle).}
\begin{center}
\includegraphics[width=13cm]{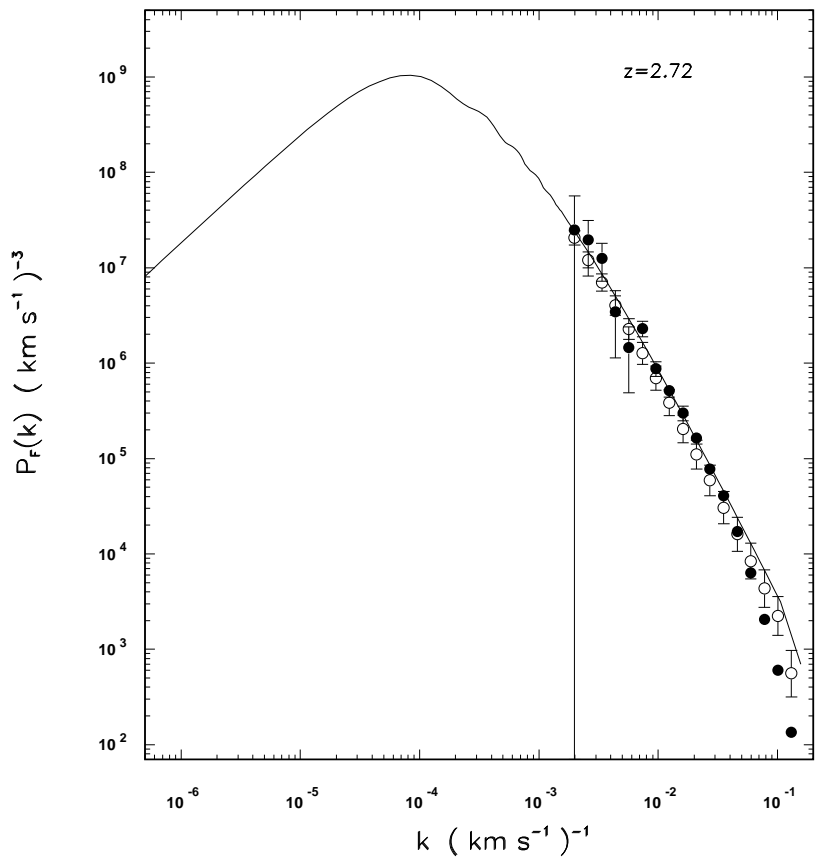}
\end{center}
\label{}
\end{figure}

\subsection{Combined CMB and Ly-$\alpha$ analysis}

By jointly considering the WMAP anisotropy data and the
Ly-$\alpha$ observations,  
we investigate the cosmological constraints
obtained on the basis of the analysis of the Ly-$\alpha$ cumulative 
mass function.
To this purpose, we use here only the accurate WMAP temperature 
anisotropy (TT) power spectrum.
In fact, although crucial to probe the cosmological reionization,
the inclusion of the polarization (ET) power spectrum 
derived from WMAP [see also the DASI detection/upper limit on
(E and B) polarization power spectrum; Kovac et al. 2002]  
does not change significantly our quantitative results,
as we have verified for a representative set of cases.
We will take into account the polarization information 
in future works.
We ran the CMBFAST code v4.2 with the COBE normalization option
to generate the CMB temperature anisotropy power spectra 
for the considered grid of parameters 
and then renormalized each computed power spectra, $C_\ell$,
of the grid to minimize the $\chi^2$ when 
compared to the WMAP data (we find that this 
renormalization does not change appreciably the final 
best fit and error bar results).
We vary $n_s$, $w$ and
$z_r$ in the same range as in the previous analysis,
 but for this case we also
 allow to vary the effective running of the slope $dn_{s}/d{\rm ln}k|_{k_p}$ at
 $k_p=0.05$Mpc$^{-1}$,
the same pivot wavenumber used 
in the analysis by the WMAP team (Spergel et al. 2003).\\
Our parameter vector has   also four dimensions:
${\bf p}=(n_s, w, z_r, dn_{s}/d{\rm ln}k$),
 where   $dn_{s}/d{\rm ln}k$
 was free to vary in the range $[-0.1 , 0.1]$ with a step of 0.01.
We compute the $\chi^2$ for each choice
of the parameters in the grid, comparing the simulated CMB temperature anisotropy
power spectra with the  WMAP anisotropy  power spectrum,
following the same procedure as in the previous analysis.
In addition, for each choice of parameters in the grid
we generate the matter transfer function and
evaluate the linear matter power spectrum $P_L(k)$ at the
present time  with the normalization given by the equation (22).
For each $P_L(k)$  we compute  the effective slope
$n_{{\rm eff}}(k_p)$ and the running of the slope $d n_{\rm eff}/d {\rm ln}k$
at the pivot wavenumber $k_p=0.05$Mpc$^{-1}$, as given by equations (24)--(26)
by only replacing $k^*_p$ with $k_p$.
Note the difference between the two definitions of the running of the slope:
$d n_s/d {\rm ln}k$ is the matter power law free parameter 
(see Spergel et al. 2003 and
the CMBFAST code v4.2) while  $d n_{\rm eff}/d {\rm ln}k$ is obtained
from the matter transfer function shape.  \\
\begin{table}[]
\caption[]{Cosmological constraints from the joint WMAP and Ly-$\alpha$
analysis.
The values of  $n_{{\rm eff}}$, $dn_{s}/d\,{\rm ln}k$ and   $dn_{\rm eff}/d\,{\rm ln}k$
are evaluated at the pivot wavenumber $k_p$=0.05Mpc$^{-1}$;
$T_{r,4}$ is the reionization temperature in units of $10^4$K.}
\begin{center}
\begin{tabular}{ccc}
\hline \hline
                               & WMAP+Ly-$\alpha$&         \\ \hline
Parameter                      &$68\%$ CL & $95\%$ CL            \\ \hline
$n_s$                          & 1.034$\pm$0.043 &1.011$\pm$0.059\\
$T_{r,4}$                      &2.320$\pm$0.194& 2.228$\pm$0.229\\
$z_r$                          &26.131$\pm$3.105&25.495$\pm$3.728\\
$w$                            &$-$0.804$\pm$0.122&$-$0.831$\pm$0.115\\
$\sigma_8$                     &0.945$\pm$0.035&0.949$\pm$0.032 \\
$\tau^a$                       &0.167$\pm$0.028&0.159$\pm$0.032 \\
$\sigma_8 e^{-\tau}$           &0.800$\pm$0.037&0.810$\pm$0.040 \\
$n_{{\rm eff}}$                &$-$2.151$\pm$0.045&$-$2.154$\pm$0.047\\
$dn_{\rm eff}/d\,{\rm ln}k$    &0.018$\pm$0.004&0.017$\pm$0.005\\
$dn_{s}/d\,{\rm ln}k$  &0.023$\pm$0.022&0.024$\pm$0.023 \\ \hline
$^a$Assumes ionization fraction, $x_e$=1.
\end{tabular}
\end{center}
\end{table}
\begin{figure}
\caption{Panel a): constraints  in  $n_{{\rm eff}}$ - $dn_s/d{\rm ln}k$
plane from  MCMC with $dn_s/d{\rm ln}k \ne 0$.
Panels b) and c): constraints in
$n_{{\rm eff}}$~--~$\sigma_8e^{-\tau}$ plane and
$\sigma_8e^{-\tau}$ - $d n_{{\rm eff}}/d {\rm ln }k$
plane from  MCMC with $dn_{\rm eff}/d{\rm ln}k =0$.
All contours and error bars are at $68\%$~CL; $n_{{\rm eff}}$,  
$dn_s/d{\rm ln}k$ and
$dn_{\rm eff}/d{\rm ln}k$ are obtained at $k_p=0.05$Mpc$^{-1}$.}
\begin{center}

\vspace{-8.3cm}
\includegraphics[width=15cm]{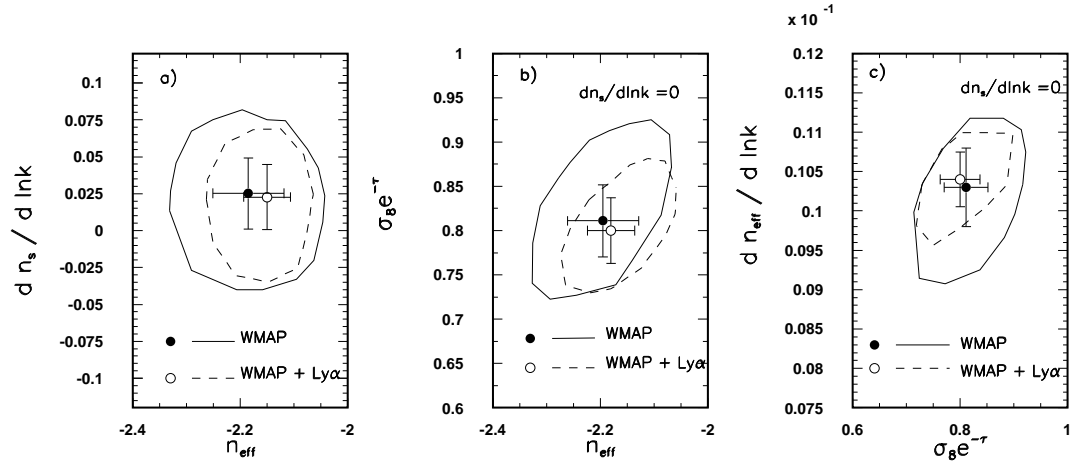}
\end{center}
\label{}
\end{figure}
We generate Monte Carlo Markov Chains (MCMC)
using the combined $\chi^2$ obtained
from the Ly-$\alpha $ cumulative mass function  and WMAP anisotropy analysis
(the number of model elements is of about $3 \times 10^6$).
In this way we sample the chi-squared probability distribution in the combined
parameter space. We run the  MCMC  with $dn_s/d{\rm ln}k=0$
and varying $dn_s/d{\rm ln}k$ in the range mentioned above.\\
Before of discussing the results obtained
by using the pivot wavenumber $k_p$=0.05Mpc$^{-1}$
we briefly report on the results we derived
by using the pivot point $k^*_p$=0.03~s/km
as in Sect.~3.1 but by exploiting only the WMAP
(TT) power spectrum or WMAP combined to the Ly-$\alpha$
information. They are shown again in Figure~3: note how
for this pivot point choice
the poor sensitivity of WMAP at the small scales
accessible to Ly-$\alpha$ observations does not
improve but slightly worses the parameter recovery,
the two kinds of observations having
similar degeneracy directions in the  $n^*_{\rm eff} - \Delta^2_*$
 plane.
On the contrary, the situation improves by considering a pivot
wavenumber (namely at $k_p$=0.05Mpc$^{-1}$) at larger scales.
We report in  Table~2 the results obtained from the joint
WMAP and Ly-$\alpha$ analysis
from the MCMC with $dn_s/d{\rm ln}k \ne 0$  that
can be compared  with the WMAP results
(Spergel et al. 2003; Peiris et al. 2003). \\
Panel a) in Figure 5 presents the constraints
in  $n_{\rm eff}$ - $dn_s/d{\rm ln}k$ plane  obtained
from the analysis of the WMAP anisotropy measurements and
the joint WMAP and Ly-$\alpha$
analysis from MCMC with  $dn_s/d{\rm ln}k \ne 0$.
We found that both analyses favour a positive running
of the slope $dn_s/d{\rm ln}k \approx 0.023$
and an effective spectral index $n_{eff}\approx-2.2$ at
$k_p$=0.05 Mpc$^{-1}$.\\
Panels b) and c)
present the constraints in
the $n_{{\rm eff}}$~--~$\sigma_8e^{-\tau}$ plane
and $\sigma_8e^{-\tau}$~--~$dn_{\rm eff}/d{\rm ln}k$ plane
 obtained from MCMC with
$dn_s/d{\rm ln}k =0$~\footnote{We have verified that
the value of $dn_{\rm eff}/d{\rm ln}k$ may depend
quite critically on the
choice of the step in $k$ used to numerically compute the
(central) derivatives of the power spectrum.
For the considered models
we find in practice quite stable results
for steps in $k$ less than $\simeq 5$~\% of $k$,
while using larger steps significantly affects
the final results.}.
From the  Monte Carlo Markov chain
with $dn_s/d{\rm ln}k =0$ we found  $n_{{\rm eff}}\approx -2.19$ and
$dn_{{\rm eff}}/d{\rm ln}k \approx 0.01$.

Our results differ from the value of the effective
running of the slope $dn_s/d{ln}k \approx  -0.03$
found by the WMAP team (Spergel et al. 2003; Peiris et al. 2003)
at the same pivot wavenumber and indicate
that a possible identification of a running of the slope,
$dn_s/d{\rm ln}k\ne 0$, at $k_p$=0.05Mpc$^{-1}$
(multipole ${\it l} \approx 700$) with the current data is
mainly an effect of the existing
degeneracy in the amplitude-slope plane at this scale, the
result being clearly consistent with the absence of running.

In Figure 6 we compare the cosmological parameter constraints at $68\%$ CL
from the Ly-$\alpha$ analysis and the joint WMAP and Ly-$\alpha$ analysis with the
cosmological parameter ``simulated'' constraints 
on cosmological parameters achievable by WMAP after 4 years 
of observations and by 
the combination of the three ``cosmological'' channels of
{\sc Planck} (Mandolesi et al. 1998, Puget et al. 1998, Tauber 2000) 
at 70, 100, and 143~GHz
considering  only the multipoles $\ell \le 1500$
and neglecting the Galactic and extragalactic foreground contamination. 
We assume Gaussian symmetric beams with the nominal resolution, 
a sky coverage of $80\%$, 
the cosmic variance and nominal noise sensitivity 
as sources of error, and neglect 
for simplicity possible systematic effects.
Only the information from the temperature (TT) power spectrum
is again considered.  
In the case of the ``simulated'' data we assume exactly
the current WMAP data and error bars at $\ell \lsim 300$, 
being the uncertainty in that 
multipole range dominated by the cosmic variance.
We consider as fiducial model the best fit power law
$\Lambda$CDM model to the WMAP data with $dn_s/d\,{\rm ln}k=0$
(Table 1 from Spergel at al. 2003).

Note the improvement 
on power spectrum and reionization parameters  
achievable by using the final WMAP data
(improvement of about a factor of two) 
and that (of about a further factor of two)
achievable with {\sc Planck} 
by using only the temperature anisotropy data. 
Note also the role of {\sc Planck} in reducing 
the error bar for the parameter $w$ defining the equation of state 
of the dark energy component
parameter.

We find that adding the current Ly-$\alpha$ information
to the simulated WMAP~4-yr data only slightly reduces 
the error bars (of course, the relative improvement is 
significantly smaller by adding them to the simulated {\sc Planck}
data).

\begin{figure}
\caption{Cosmological parameter constraints at $68\%$ CL
from the Ly-$\alpha$ and  
WMAP+Ly-$\alpha$ analysis compared with the
cosmological parameter constraints achievable 
by WMAP after 4 years of observations and by 
{\sc Planck}. For all panels the meaning of the symbols is 
the same as in panel a). See also the text.} 
\begin{center}
\includegraphics[width=15cm]{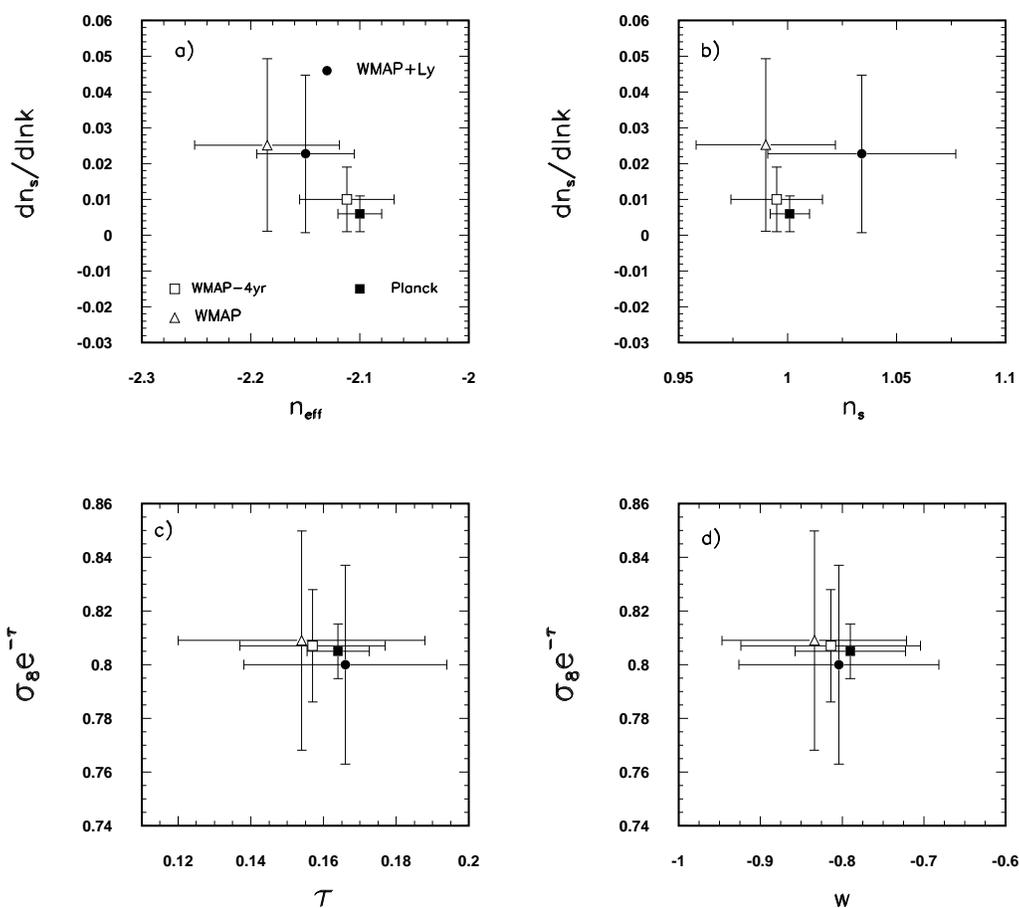}
\end{center}
\label{}
\end{figure}

\section{Discussion and conclusions}

The recent detection of high values of the electron optical depth
to the last scattering (Kogut et al. 2003, Spergel et al. 2003) by the WMAP
satellite (Bennett et al. 2003) implies the existence of  an early epoch of
reionization of the Universe at $z_r\sim 20$, fundamentally
important for  understanding  the formation and evolution of the
structures in the Universe. \\
As the reionization is assumed to be caused by the ionizing photons produced
during the early stages of star-forming galaxies and quasars, we evaluate this effect by computing
the cumulative mass function of the high-z bound objects
for a class of flat cosmological models with cold dark matter plus cosmological constant
or dark energy with constant equation of state, encompassing different 
reionization scenarios.
Our fiducial cosmological model has
$\Omega_bh^2=0.024\pm 0.001$, $\Omega_mh^2=0.14\pm0.02$, $h=0.72\pm0.05$
as indicated  by the best fit power law $\Lambda$CDM model of WMAP data
(Spergel et al. 2003).

Assuming that the virilization takes place at the collapse time
and a constant baryon/dark matter ratio in collapsed objects,
we compute the fraction of the mass residing in gravitationally
bounded systems as a function of the redshift of collapse at each linear scale
and of the virial mass-temperature relation.
We evaluate the formation rate of bound objects at $z=2.72$
and their cumulative mass function was compared with the
cumulative mass function obtained from the 
Ly-$\alpha$ transmission power spectrum (Croft et al. 2002). \\
Our method allows to study reionization models
by varying the amplitude, spectrum, and epoch of the reionization and the
cosmological parameters.
In the same time, as the high-z bound objects
are rare fluctuations of the overdensity field, the tail of the cumulative mass
function is sensitive to the {\it rms} mass
fluctuations within the filtering scale $\sigma(R_f,z)$.\\
 We find that the analysis of the cumulative mass function of
 the Ly-$\alpha$ systems indicates a reionization redshift 
 in agreement with the value found on the basis of the WMAP
 anisotropy measurements,
 setting constraints on the amplitude of
 the power spectrum, $\sigma_8$,
 similar to those derived
from the X-ray cluster temperature function.

Our joint analysis of  Ly-$\alpha$ cumulative mass function
and WMAP anisotropy measurements
shows that a possible identification of a running of the slope,
$dn_s/d{\rm ln}k\ne 0$, at $k_p$=0.05Mpc$^{-1}$
(multipole ${\it l} \approx 700$) is
mainly an effect of the existing
degeneracy in the amplitude-slope plane at this scale, the
result being clearly consistent with the absence of running,
the other constraints based on WMAP remaining substantially
unchanged.

We also shown that, for the set of cosmological models
studied in this work, the error bars on the considered parameters
can be reduced by about a factor of two by using the
final WMAP data.
The temperature anisotropy data
from the forthcoming {\sc Planck} satellite
will further improve the sensitivity on these parameters,
by another factor of two, and also
the reliability of these results thanks to the better
foreground subtraction achievable with the wider frequency
coverage and the improved sensitivity, resolution and
systematic effect control.
This information jointed 
with the great improvement on the study 
of the Ly-$\alpha$ forest trasmission
power spectrum (Seljak et al. 2002)
achievable by the increase of the number of 
quasar spectrum measures 
expected from the Sloan Digital Sky Survey 
will allow to significantly better constrain
the properties of the primordial density field
at small scales.

 \section{Acknowledgements}

We acknowledge the use of the computing system at
{\sc Planck}-LFI Data Processing Center in Trieste
and the staff working there.
LAP  acknowledge  the financial support from the European
Space Agency.
It is a pleasure to thank to U. Seljak and M. Zaldarriaga
for the use of the CMBFAST code v4.2 employed in the computation
of the CMB power spectra and the matter transfer functions.

\newpage

\end{document}